# Experimental Demonstration of Signal-to-Noise-Ratio Improvement of Fourier-Domain Optical Coherence Tomography

**Paul Blazkiewicz, P. Malcolm Gourlay**

Centre for Biophotonics and Laser Science, Physics, The University of Queensland, Brisbane, QLD

4072. Phone: + 61 7 3365 3430, Fax: + 61 7 3365 1242,

Email: *blazkie@physics.uq.edu.au*

**John R. Tucker, Aleksandar D. Rakic**

School of Information Technology and Electrical Engineering,

The University of Queensland, Brisbane, QLD 4072.

**Andrei V. Zvyagin,**

Centre for Biophotonics and Laser Science, Physics/

School of Information Technology and Electrical Engineering,

The University of Queensland, Brisbane, QLD 4072

## Abstract

A recent advance in optical coherence tomography (OCT), termed swept-source OCT, is generalized into a new technique, Fourier-domain OCT. It represents a realization of a full-field OCT system in place of the conventional serial image acquisition in transverse directions typically implemented in "flying-spot" mode. To realize the full-field image acquisition, a Fourier holography system illuminated with a swept-source is employed instead of a Michelson interferometer commonly used in OCT. Fourier-domain OCT offers a new leap in signal-to-noise ratio improvement, as compared to flying-spot OCT systems. This paper presents experimental evidence that the signal-to-noise ratio of this new technique is indeed improved.

*OCIS: (170.4500), (170.3890), (030.4280), (110.4280)*

Optical coherence tomography (OCT) is playing an important role in the advancement of non-invasive optical imaging in biomedicine by way of its ability to generate high-resolution images of unprecedented quality, deep within turbid media.[1,2] Its unique imaging advantage relies on the coherence gating which can discriminate reflections at a precise depth within the medium. The technique is based upon the Michelson interferometer with the sample located in one arm and a reference mirror in the other. The generation of interference fringes is confined to reflections occurring only at depths in the sample medium corresponding to the reference arm delay to within the coherence length of the source. By using a broadband optical source, such as a super-radiant light emitting diode, this depth resolution is on the scale of the micrometre-scale temporal coherence length, hence the term coherence gating.[3,1] In its conventional form, time-domain OCT (TD-OCT),[4] the reference optical delay is scanned to acquire one-dimensional reflectivity profile of the sample, termed A-scan. This scanning delay is typically realized by a mechanically driven actuator.[5] If the probe beam is scanned across the sample in one plane (B-scan), a series of slices, i.e., tomographs of the reflectivity profile of a sample may be built up, resulting in a two-dimensional (2D), cross-sectional morphological image of the sample. With addition of another transverse scanning plane (C-scan), a three-dimensional image of the sample is built. The rastering across the sample in the transverse directions is realized by means of "flying-spot" configuration, typically implemented by mechanical scanners. A major shortcoming of this serial pixel-by-pixel image acquisition technique is its theoretical suboptimal signal-to-noise-ratio (SNR) resulting from shot noise arising from the intense reference beam impinging on a single detector.

A novel principle to increase the SNR called spectral-domain OCT (SD-OCT) was proposed by Fercher *et al.*[6], Häusler *et al.*[7] and demonstrated by several groups.[8,9,10,11] Its principle innovation rests in optically dispersing the interferometric signal with a spectrometer. The spectral components are sampled with a linear detector array, in place of the single detector used in TD-OCT. The reference arm delay is stationary, as opposed to the scanning optical delay in TD-OCT. As there are no time-varying parameters, the optical output is constant and is integrated over exposure time. The detector array samples a series of spectral bins, which are equally spaced in wavenumber.

The axial image may be recovered by performing a digital Fourier transform (DFT) of the spectrometer output. The key advantage of SD-OCT is that the $N$ detector signals are added coherently, whereas the noise components are summed incoherently. This produces an improvement of SNR which is proportional to number of detectors and depends on the relative bandwidth of an individual detector.[12] With this approach a 21.7-dB improvement in SNR has been achieved.[10] In addition, the need for the mechanical scanning delay line is no longer necessary. This also holds true for swept-source OCT (SS-OCT), where the combination of a swept-source and single detector is used. The swept-source has a narrow linewidth and is swept over a broad wavelength range.

In spite of these advances, SD-OCT and SS-OCT still rely on serial pixel-by-pixel flying-spot image acquisition in the transverse dimensions. Apart from the inconvenience of its typical mechanical implementation, this single point sampling in the spatial dimension degrades the SNR than what could be ultimately achieved by full-field image acquisition by a parallel array.

We propose to extend the technique of image acquisition in the Fourier domain to the transverse spatial dimension, resulting in a fully-optimised technique termed Fourier-domain OCT (FD-OCT).[13] We are departing from the conventional Michelson interferometer configuration towards digital holographic techniques.[14] Our proposal may be considered as a combination of SS-OCT and digital Fourier holography. The full-field *en face* image is captured holographically in the Fourier-domain using an array in place of the conventional flying-spot technique, as shown in Fig. 1. For each swept-source wavenumber $\{k_1, k_2, \ldots, k_M\}$, a 2D-Fourier-hologram is recorded, a complete set of which forms a 3D data array. Given that each hologram is scaled by the Fourier-transformed depth information of the SS-OCT, the complete data set is essentially a 3D-Fourier-transform of the object. Hence the 3D morphological image may be reconstructed via a 3D-DFT.

We expect a further improvement in SNR from this holographic method in each of the transverse dimensions which represents a significant advance. SNR improvement is calculated to be 27 dB per transverse dimension for a typical CCD array of linear size of 1000 pixels, provided shot noise is the dominant source of noise. In addition, the sweep-source update rate is expected to be reduced to the video frame rate, in comparison with the kilohertz-rates used for the flying-spot SS-OCT.[15] In this letter, we present a crucial test-case, demonstrating experimentally that the principles of optimising the SNR extend to the transverse dimension in transmission mode. We use the transmission mode to simplify the experimental setup but it is possible to reconfigure the experiment in the reflection mode. It is important to note that the SNR improvement in context of SD-OCT was first demonstrated by de Boer *et al.*[12] using hybrid timed-domain SD-OCT (TD/SD-OCT). The demonstrated improvement was subsequently confirmed by greatly improved imaging performance of SD-OCT systems. In close analogy with this work, we present the demonstration of the SNR improvement in one transverse dimension using hybrid TD/FD-OCT.

The experimental setup of the hybrid TD/FD-OCT system is presented in Fig. 2. The optical circuit represents a Fourier holography system with one important difference: the reference fiber is not stationary. An output beam from He-Ne laser of wavelength $\lambda = 632.8$ nm is equally split (50:50) between the fiber-optic sample and reference arms, whose distal ends are situated at the front focal plane of a Fourier lens of the focal length $f = 100$ mm. The fibers are closely separated in the vertical $y$-direction ($< 1$ mm) to allow the horizontal translation of the reference fiber, i.e. along the $x$-axis (see Fig. 2). The sample fiber is placed on-axis and represents a point-like source. The reference fiber produces a plane reference wave that is incident at an angle on the detector that is placed at the back focal plane of the Fourier lens. The reference and sample optical fields interfere across the detector array to produce an interference pattern, assuming equal power in both arms the irradiance, $I$, is given by the following equation:

$$I = 4I_o \cos^2\left[\left(\frac{\boldsymbol{p}\Delta x}{\boldsymbol{\lambda} f}\right)x_{\text{det}}\right],\qquad(1)$$

where $I_0$ is the incident irradiance from one arm, $\Delta x$ denotes the separation between the reference and sample fibers, $x_{\text{det}}$ is the horizontal position of the detector array. Terms that give rise to the constant background are omitted from Eq. (1). When scanning the reference arm, $\Delta x$ varies linearly in time as $\Delta x = vt$, where $v$ is the linear translation stage velocity and $t$ is time. The formal structure of Eq.(1) is conformal to the structure of expressions obtained from Eq.(10) in the paper by de Boer *et al.*[12] in the context of hybrid SD-OCT. In TD/SD-OCT, wavenumber represents the time varying parameter, whereas in TD/FD-OCT, $\Delta x$ is the time varying parameter. A horizontal (x-direction) slit was placed in front of the detector array to eliminate the signal acquisition artefact due to the finite vertical separation of the fibers, which was necessary to allow the reference fiber non-obstructed pass across the optical axis.

Translation of the reference fiber causes variation of the fringe frequency in time. It results in the generation of a time-varying signal at each detector element, $i_d$ that is found by integration of the detector photoresponse [Eq. (1)] over the detector element width, $d$:

$$ i_d = \left( \frac{2I_o A d \eta e \lambda}{hc} \right) \left[ 1 + \cos\left( \frac{\pi v^2}{\lambda f} \right) t^2 \cos\left( \frac{2\pi n dv}{\lambda f} \right) t \, \mathrm{sinc}\left( \frac{\pi dv}{\lambda f} \right) t \right], \qquad (2) $$

where $i_d$ is the detector current, $A$ is the detector area, $\eta$ is the quantum efficiency, $e$ is the charge of an electron, $h$ is Planck's constant, $c$ is the speed of light, and $n$ is the index of the detector element, where $n = 0$ denotes the on-axis detector element. Examination of Eq. (2) shows that the signal represents a sinc-function envelope on a constant background that peaks at $t = 0$, whereas the signal carrier frequency is primarily determined by the index of the detector element; the greater the index, the greater the carrier frequency. The first factor of the a.c. term of Eq. (2) describes the frequency chirp. The chirp is not significant as long as the signal is recorded in the vicinity of the central peak of the sinc-function, and is, therefore, omitted in our analysis.

In our experiment, we individually recorded the temporal signals from four adjacent detectors for a single scan of the reference arm. The detector elements were placed considerably off-axis to attain higher carrier frequency, since the translation stage velocity was limited to 300 µm/s. The reference fiber was scanned and the time-varying signal of each detector element was digitized and stored on a personal computer.

We process the recorded four signals by using two following filtering schemes. The first scheme is representative of TD-OCT. The four channel amplitude spectrums were summed up following broad bandpass filtering using an $4^{th}$-order Butterworth filter (bandwidth 25.8 Hz). The processed signal is shown in Fig. 3 (dotted curve). The second scheme is representative of the TD/FD-OCT scheme, where each detector element time-varying signal was individually narrow band-pass filtered (bandwidth of 17.3 Hz), and these frequency responses were summed in Fourier domain. The result is shown in Fig. 3 (solid curve). The in-band noise power density in the narrow band-passed case was reduced by a factor of 3.38 and the signal power density was decreased by a factor of 1.20 improving the overall SNR by a factor of 2.82. It agrees well with the expected SNR improvement of 2.98 previously demonstrated TD/SD-OCT case[12].

In conclusion, we have introduced a new OCT technique that represents a combination of swept-source OCT and digital Fourier holography. We believe it is capable of superseding the already achieved signal-to-noise improvements demonstrated by spectral-domain OCT by utilizing a second dimension in our detection arm to further increase our SNR. In addition, Fourier-domain OCT will considerably reduce the required update rate of the swept-source as the need for raster scanning in the sample arm is eliminated. We have used a hybrid of FD-OCT and time-domain OCT techniques to test the signal-to-noise ratio improvement in the transverse dimension. Our results demonstrate that optimization of the SNR can be, indeed, generalized to include the transverse dimension.

**References (with titles)**

1. D. Huang, E. A. Swanson, C. P. Lin, J. S. Schuman, W. G. Stinson, W. Chang, M. R. Hee, T. Flotte, K. Gregory, C. A. Puliafito, and J. G. Fujimoto, "Optical coherence tomography", Science, Vol. 254, (1991), pp.1178-1181.

2. Brett E. Bouma and Guillermo J. Tearney, "Handbook of Optical Coherence Tomography", (Marcel Dekker, Inc, NewYork, 2002) pp. 26.

3. A. F. Fercher, W. Drexler, C. K. Hitzenberger "Optical coherence tomography – principles and applications" Rep. Prog. Phys., Vol.66, (2003), pp.239-303.

4. J. M. Schmitt "Optical Coherence tomography (OCT): A Review" IEEE J. Selec. Top. Quant. Electron., Vol.5, (1999), pp.1205-1215.

5. M. E. Brezinski and J. G. Fujimoto "Optical Coherence Tomography: High Resolution Imaging in Nontransparent Tissue" IEEE J. Selec. Top. Quant. Electron., Vol.5, (1999), pp.1185-1192.

6. A. F. Fercher, C. K. Hitzenberger, G. Kamp, S.Y. Elzaiat "Measurement of intraocular distances by backscattering spectral interferometry" Opt. Comm., Vol. 117, (1995), pp.43-48.

7. G. Häusler and M. W. Lindner ""Coherence Radar" and "Spectral Radar" – new Tools for Dermatological Diagnosis" J. Biomed. Opt. **3** (1998) pp.23-31.

8. M. Wojtkowski, R. Lietgeb, A. Kowalczyk and T. Bajraszewski "Real-time in vivo imaging by high-speed spectral optical coherence tomography" Opt. Lett., Vol. 28, (2003), pp.1745-1747.

9. M. A. Choma, M. V. Sarunic, C. Yang, J. A. Izatt "Sensitivity advantage of swept source and Fourier domain optical coherence tomography" Opt. Expr., Vol. 11, (2003), pp.2183-2189.

10. N. Nassif, B. Cense, B. H. Park and S. H. Yun, T. C. Chen, B. E. Bouma, G. J. Tearney and Johannes F. de Boer " In vivo human retinal imaging by ultrahigh-speed spectral-domain optical coherence tomography" Opt. Lett., Vol.29, (2004), pp.480-482.

11. R. Leitgeb, C. K. Hitzenberger, A. F. Fercher, "Performance of fourier domain vs. time domain optical coherence tomography", Opt. Expr., Vol.11, (2003), pp.889-894.

12. J. F. de Boer, B. Cense, B. H. Park, M. C. Pierce, G. J. Tearney and B. E. Bouma "Improved signal-to-noise ratio in spectral-domain compared with time-domain optical coherence tomography" Opt. Lett., Vol.28, (2003), pp.2067-2069.

13. A. V. Zvyagin, "Fourier-Domain Optical Coherence Tomography: Optimization of Signal-to-Noise Ration in Full Space", submitted to Optics Communications.

14. E. Wolf, "Three dimensional structure determination of semi-transparent objects from holographic data", Opt. Comm. Vol.1, No. 4, (1969) pp.153-156,

15. S. H. Yun, G. J. Tearney, J. F. de Boer, N. Iftimia and B. E. Bouma, "High speed optical frequency-domain imaging", Opt. Expr. Vol. 11, No. 22, (2003), pp.2953-2960

**References (without titles)**

1. D. Huang, E. A. Swanson, C. P. Lin, J. S. Schuman, W. G. Stinson, W. Chang, M. R. Hee, T. Flotte, K. Gregory, C. A. Puliafito, and J. G. Fujimoto, Science, Vol. 254, (1991), pp.1178-1181.

2. Brett E. Bouma and Guillermo J. Tearney, "Handbook of Optical Coherence Tomography", (Marcel Dekker, Inc, NewYork, 2002) pp. 26.

3. A.F. Fercher, W. Drexler, C. K. Hitzenberger,  Rep. Prog. Phys. Vol. 66 (2003) pp.239-303.

4. J. M. Schmitt, IEEE J. Selec. Top. Quant. Electron. Vol. 5 (1999) pp.1205-1215.

5. M. E. Brezinski and J. G. Fujimoto, IEEE J. Selec. Top. Quant. Electron., Vol.5, (1999), pp.1185-1192.

6. A.F. Fercher, C. K. Hitzenberger, G. Kamp, S.Y. Elzaiat, Opt. Comm., Vol. 117, (1995), pp.43-48.

7. G. Häusler and M. W. Lindner, J. Biomed. Opt., Vol.3, (1998), pp.23-31.

8. M. Wojtkowski, R. Lietgeb, A. Kowalczyk and T. Bajraszewski, Opt. Lett., Vol.28, (2003), pp.1745-1747.

9. M. A. Choma, M. V. Sarunic, C. Yang, J. A. Izatt, Opt. Expr., Vol. 11, (2003), pp.2183-2189.

10. N. Nassif, B. Cense, B. H. Park and S. H. Yun, T. C. Chen, B. E. Bouma, G. J. Tearney and Johannes F. de Boer, Opt. Lett., Vol. 29, (2004), pp.480-482.

11. R. Leitgeb, C. K. Hitzenberger, A. F. Fercher, Opt. Expr. 11 (2003) pp.889-894.

12. J. F. de Boer, B. Cense, B. H. Park, M. C. Pierce, G. J. Tearney and B. E. Bouma, Opt. Lett., Vol. 28, (2003), pp.2067-2069.

13. A.V. Zvyagin, (submitted to Optics Communications).

14. E. Wolf, Opt. Comm. Vol.1, No. 4, (1969) pp.153-156,

15. S. H. Yun, G. J. Tearney, J. F. de Boer, N. Iftimia and B. E. Bouma, Opt. Expr. Vol. 11, No. 22, (2003), pp.2953-2960

**Figure 1**     Schematic diagram of FD-OCT.  C, collimator; ODL, optical delay line.  The reference fiber is stationary.

**Figure 2**     Schematic diagram of the hybrid TD/FD-OCT experimental setup used for the test case of the SNR-improvement in the transverse direction.  S, sample fiber; R, horizontally ($x$-axis) scanned reference fiber; TCA, trans-conductance amplifier.  In Digital sensor, shaded boxes represent schematically off-axis detectors used to record the time-varying signal.

**Figure 3**     Logarithmic plot of a signal acquired from four detectors using experimental setup of Fig. 2.  In the first case (dotted curve), the signal components are summed up following broad bandpass-filtering, which is equivalent to the signal processing of time-domain OCT.  In the second case (solid curve), the signal components are narrow bandpass-filtered and then summed up, which is equivalent to the hybrid TD/FD-OCT, and, ultimately relevant to the SNR improvement of FD-OCT.

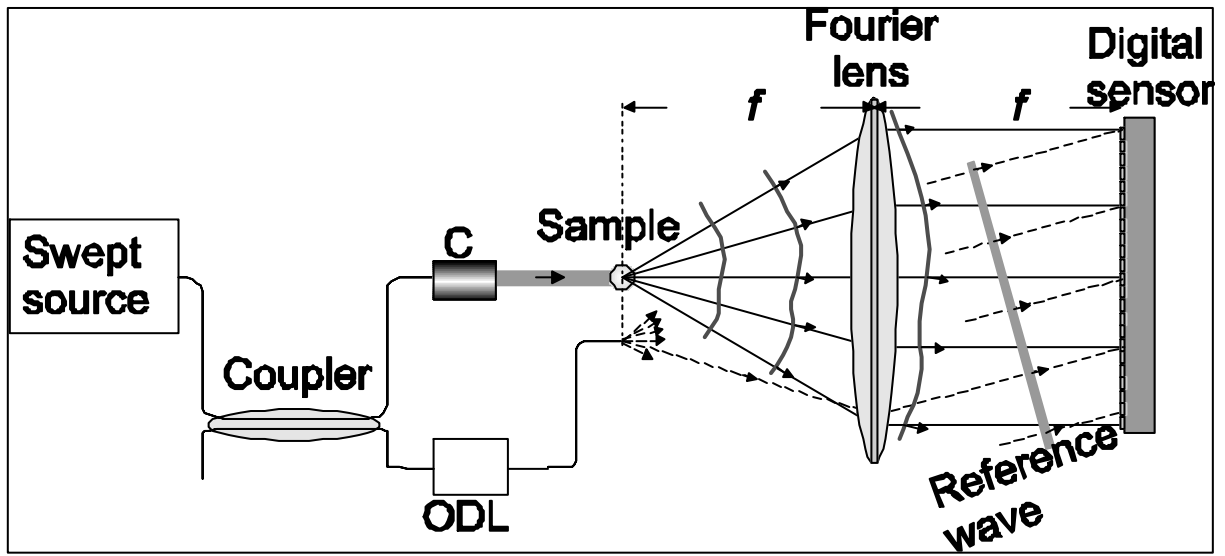

**Figure 1**

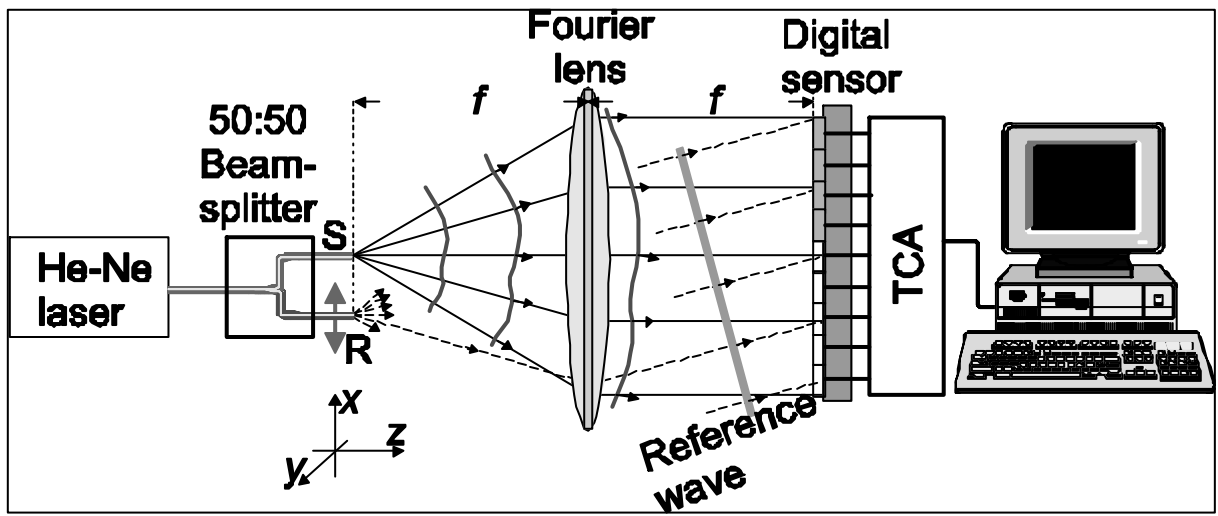

**Figure 2**

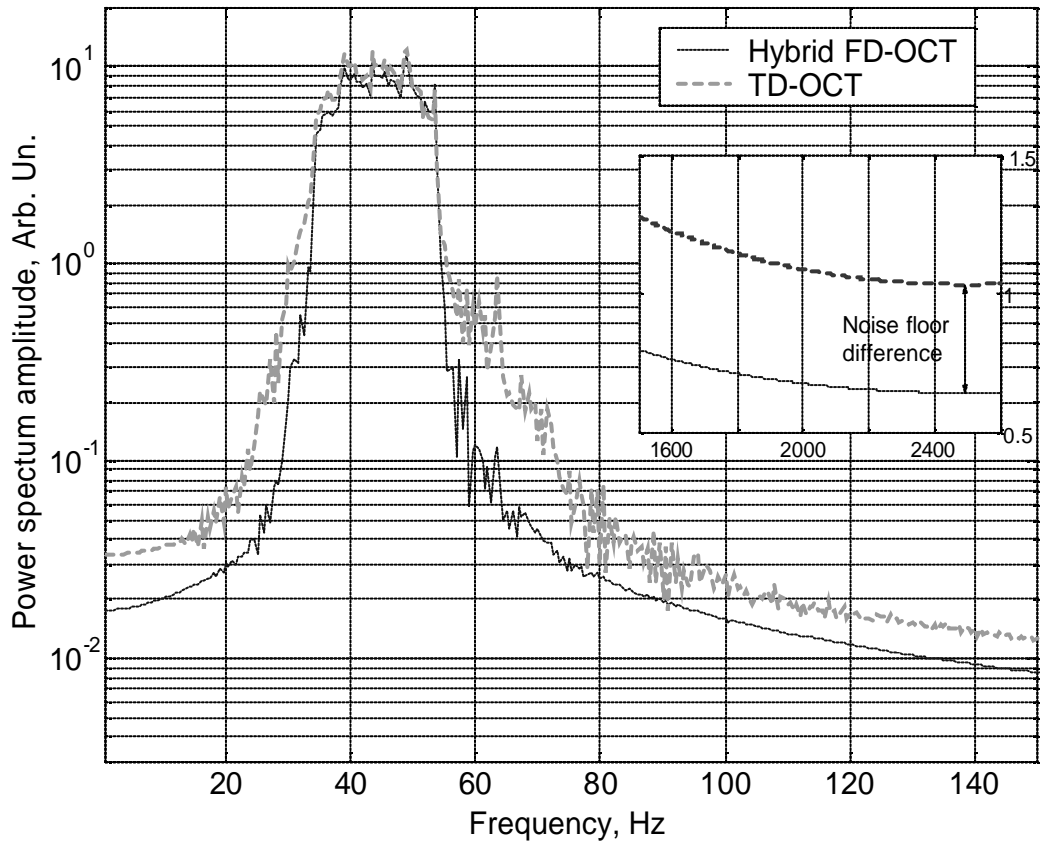

**Figure 3**